\documentstyle[epsf,12pt]{article}
\newcommand{\fg}[2]{\null\centerline{\epsfxsize=#1truecm\epsffile{#2}} 
}

\newcommand{\beq}{\begin{equation}}
\newcommand{\bfg}[1]{\begin{figure}[#1]}
\newcommand{\efg}{\end{figure}}
\newcommand{\eeq}{\end{equation}}
\newcommand{\brry}{\begin{eqnarray}}
\newcommand{\rc}{\nonumber\\}
\newcommand{\erry}{\end{eqnarray}}

\newcommand{ \eq}{\ =\ }
\newcommand{\moy}[1]{\mbox{$\langle #1 \rangle $}}
\newcommand{ \h}{{\bf h}}

\newcommand{\clDq}{{\cal D}_q}
\newcommand{\clFq}{{\cal F}_q}

\newcommand{\expp}[1]{\exp{\big(#1\big)}}
\newcommand{\as}{\alpha_s}

\newcommand{\go}{\gamma_0}

\begin{document}

\title{
 GENERALISED FACTORIAL MOMENTS \\
 AND QCD JETS\thanks{Preprint Nice INLN 94/6 }}

\author{ P. Duclos and J.-L. Meunier\thanks
{I.N.L.N., Universit\'e
de Nice-Sophia
Antipolis, Unit\'e Mixte
de Recherche du CNRS, UMR 129, 1361 Rt. des Lucioles 06560 Valbonne,  
France:}}
\maketitle
\begin{abstract}
{
In this paper we present a natural and comprehensive generalisation  
of the standard factorial moments ($\clFq$) analysis of a  
multiplicity distribution. The Generalised Factorial Moments  are  
defined for all $q$ in the complex plane and, as far as the negative  
part of its spectrum is concerned, could be useful for the study of  
infrared structure of the Strong Interactions Theory of high energy  
interactions (LEP multiplicity distribution under the ${\cal Z}_0$).  
The QCD calculation of the Generalised Factorial Moments   for  
negative  $q$ is performed in the double leading log accuracy and is   
compared to OPAL experimental data. The role played by the infrared  
cut-off of the model is discussed and illustrated with a Monte Carlo  
calculation.  }
\end{abstract}
\vskip 1truecm

\section{Introduction}
%%%%%%%%%%%%%%%%%%%%%%%%%%%%%%
In this paper we present a natural generalisation of the standard  
factorial moments to continuous or fractional orders  of a  
multiplicity distribution, which could be of interest in the study of  
the infrared structure of the strong interaction theory.

In the past year, three groups of authors\cite{BMP,DD93,OW93} have  
shown that it was possible to define and compute a multi fractal  
dimension, $\clDq$, for QCD. Technically, this has been possible by  
computing the positive (and integer) order Factorial Moments of the  
distribution of particles in a restricted open angle $\Delta$ and  
could be compared with, say, the charged particle distribution in the  
${\cal Z}_0$ decay at LEP\cite{LEP}.
\begin{equation}
\label{inter}
{\cal F}_q(\Delta)\equiv {\moy{n(n-1)..(n-q+1)}_\Delta \over  
\moy{n}_\Delta^q} \propto \Delta^{(q-1)(1-{\cal D}_q/d)} 
\end{equation}
where d is the dimension of the phase space under consideration  
($d=2$ for the
whole angular phase space, and $d=1$ if one has integrated over, say  
the
azimutal angle).

In the constant coupling case $\clDq$ is well defined and
reads: 

\begin{equation}
\label{diminter}
{\cal D}_q\eq\go\frac{q+1}{q} 
\end{equation}
where $\go^2=4C_A{\as/ 2\pi}$, $\as$ is the strong interaction  
coupling constant, $C_A$ is the gluon color factor.

Dimensions ${\cal D}_q$ are called Fractal
because they come from the natural generalisation  to discrete  
variables of the
standard moments which are used in the multifractal analysis of a  
continuous
variable\cite{Generev}.
However, in this  last field, the index $q$ range is the whole
real axis, while in our case  it is restricted to positive integers.

The choice of the factorial moments as a specific tool for the study  
of the scaling behaviour of the high energy multiplicity  
distributions have been of importance. As a matter of fact, it has  
been noticed by A.Bialas and R.Peschanski\cite{BP} that the use of  
this observable permits to extract the dynamical signal from the  
Poisson noise in the Intermittency analysis of the multiplicity  
signal in high energy reactions. At first sight, the factorial  
moments ${\clFq}$ are only defined on integer and positive values of  
$q$ and does not gives any insight on the negative part of the  
multifractal spectrum (if any) of the nuclear matter.

This has been noticed some years before by R.Hwa\cite{HWA1} who first  
proposed a multifractal analysis of the signal by means of the  
so-called G-moments. However those moments did not have the property  
of the factorial moments to disentangle the Poisson noise from the  
dynamical signal, and thus suffer from statistical uncertainties.  
Further works are in progress in this direction\cite{HWA2}.

>From another point of view, one can
understand, through their definition,  that the standard Factorial  
Moments of the
 distribution are sensitive to the occurrence in the distribution of    
rare events of very high values
 of $n$ as compared to its   mean value $n_b$. For example, this is  
why the
 NA22 event\cite{NA22} has been so important in the discovery of the  
intermittent
 properties of the high  energy data.

In contrast, the negative part of the
$q$  multifractal  spectrum focusses itself on the study of rare  
events of
 relatively low values of the studied variable, which corresponds in  
our case to
 low multiplicity  events (as compared to the mean value of the  
variable). The
 moments presented in the following  have this property and are a  
natural and
 non trivial generalisation of the standard  ones.

However, in order to be
 efficient, one has to work with a multiplicity distribution with  
relatively
 high mean value, $n<<n_b$. This is why
we will not apply this analysis to the intermittency analysis of the  
data  but to the global
 multiplicity distribution and to its scaling properties with respect  
to the energy.

Let us recall that the mean particle multiplicity
 produced by a gluon of energy E disintegrating in a cone of opening  
angle
 $\Theta_0$ is given, in the Double Leading-log Approximation (DLA),
 by\cite{QCD}:
 \begin{equation}
\label{nb}
  n_b\propto \bigl[{E\Theta_0\over  \mu}\bigr]^{\go} 
\end{equation} 

were $\mu$ is the infrared cut-off of the theory, and that the  
corresponding global
  standard Factorial Moments follow  by the KNO\cite{KNO}  
phenomenon~: 

\beq 
F_q=\moy{n(n-1)..(n-q+1)}=c_q n_b^q 
\eeq
where the $c_q$ are known constants.  At first sight, it could be   
difficult to  

understand that one could find out some scaling properties from  
moments which scales with $n_b$. However, as it will be  clear in the  
following, the Generalised Factorial Moments (GFM) analysis will show  
up a non trivial behaviour with energy.

On the other hand  it is known that the
standard QCD factorial moments calculated in the DLA aproximation do  
not
fit correctly the experimental data\cite{OPAL} and that important  
corrections are
needed in order to describe the experiment reasonably . Some progress  
has been
recently made in this direction\cite{D93}.  In this paper we will  
however restrict
ourselves to the  DLA  approximation of the theory for the  
calculation of the
GFM of negative arguments of QCD  and we will study the effect of the  
infrared cut-off of the theory on this observable. We will work both  
analytically and numerically with a simple QCD
Monte-Carlo model (based on the Fragmentation  
structure\cite{BMP,MP,GMP}). We also restrict ourself to the fixed  
coupling constant case.  In all this calculation, we did not  
introduce any ad-hoc parameter exept the perturbative coupling  
constant, $\go$ which we fixe around .5 at the ${\cal Z}_0$.

As a consequence of these restrictions, we did not try to make any  
precise comparison with the experimental data.  We leave
 to further works a more precise study of the subleading
QCD corrections, running coupling constant and/or non  perturbative  
corrections which may be important in this
specific field.

The paper is organised as follows : In section 2 we
present the Generalised Factorial  Moments (GFM) and their behaviour  
for some
useful examples such as Poisson distribution,   self-similar  
structure (KNO)
or Negative Binomial distributions. Then, in section 3, we discuss  
the
Generalized Factorial Moments of QCD in the Double Leading-log  
Approximation. Then we discuss the introduction of the infrared  
cut-off in the theory. We conclude in
section 4.
%%%%%%%%%%%%%%%%%%%%%%%%%%%%%%%%%%%%%%%%%%%%%%%%%%%%%%%%%%%%%%%%%%%%%% 
%%%%%%%%%%%%%%%%
\section{The Generalised Factorial Moments.}
\subsection{ Definition.}
The standard Factorial Moments of a multiplicity distribution $P_n$  
are given by~:
\brry
\label{stafac}
F_q &\eq &\moy{n(n-1)(n-2)(\dots)(n-q+1)}_P \rc
    &\eq & \sum_0^\infty P_n \ n(n-1)(\dots)(n-q+1)
\erry
which, using the properties of the $\Gamma$ (Euler) function can be  
writen as~:
\beq
\label{GFM}
F_q=\sum_0^\infty P_n \frac{n!}{\Gamma(n-q+1)}
\eeq
and under this form can be continued in the complex $q$ plane .

The importance of the Factorial Moments comes mainly from the fact  
that they can be derived from the generating function $G(z)$; from  
the theoretical point of view it is in general much easier to handle  
than the multiplicity itself. One has :
\beq
\label{Gdef}
F_q=\frac{\partial^q\ G(z)}{\partial z^q}\Big|_{z=1},\ \ \  
G(z)=\sum_0^\infty z^n \ P_n
\eeq
which one has to generalise to continuous or fractional values of  
$q$.

Let us recall here the properties of the principal value distribution  
$x^{-q-1}$ defined on $[0,\infty[$ with respect to the convolution of  
functions defined on the positive real axis (causal  
functions)\cite{GCh}:
\beq
\label{convol}
{x^{-q-1}*f(x)\over \Gamma(-q)}\eq\int_0^x {t^{-q-1}f(x-t)\over  
\Gamma(-q)}\eq\partial_q f(x)
\eeq
where $f$ is any well behaved function, continuous and indefinitely  
differentiable at x=0. When the exponent $q$ is positive, the action  
of $x^{-q-1}$ (principal value) on a test function $\varphi$ is given  
by~:
\beq
\label{pp}
(x^{-q-1}, \varphi(x))=\int_0^\infty  
x^{-q-1}\Biggr(\varphi(x)-\Bigl\{\varphi(0)+x\varphi'(0)+\dots+{\varphi^{(n_q)}(0)\over n!}\Bigr\}\Biggr),\ \ n_q={\rm Int}(q)
\eeq
where ${\rm Int}(q)$ is the integer part of the real part of $q$.

When the real part of $q$ is negative, the integral is defined and  
can be calculated. When $q$ is positive or 0, the principal value  
ansatz (\ref{pp}) must be used, and, on positives integers the result  
is just the $q^{th}$ derivative of the function. This comes from the  
fact that for $q$ integer and positive the integral diverges together  
with the $\Gamma$ function at the denominator of \ref{convol}.

Using this definition of the derivative one can define~:
\beq
\label{cder}
F_q=\partial_q  
G(z)\Big|_{z=1}\eq{1\over\Gamma(-q)}\int_0^1G(1-t)t^{-q-1}dt
\eeq
In order to be complete, one has to prove that this definition is  
consistent with that of Eq. (\ref{Gdef}). It is easy to verify that  
introducing the definition of the generating function $G$ in Eq.  
(\ref{cder}) one recovers Eq. (\ref{GFM}) thanks to the property of  
the ($B$) Euler function of the second kind.
\subsection{Examples.}
\paragraph{\it i) The Poisson case.} In this case, $P_n$ is given by  
:
\beq
P_n={n_b^n\over n!}\expp{-n_b},\ \ \ G(z)=\expp{n_b(z-1)}
\eeq
where $n_b$ is the mean value of $n$, and it is easy to obtain :
\bfg{thb}
\fg{15}{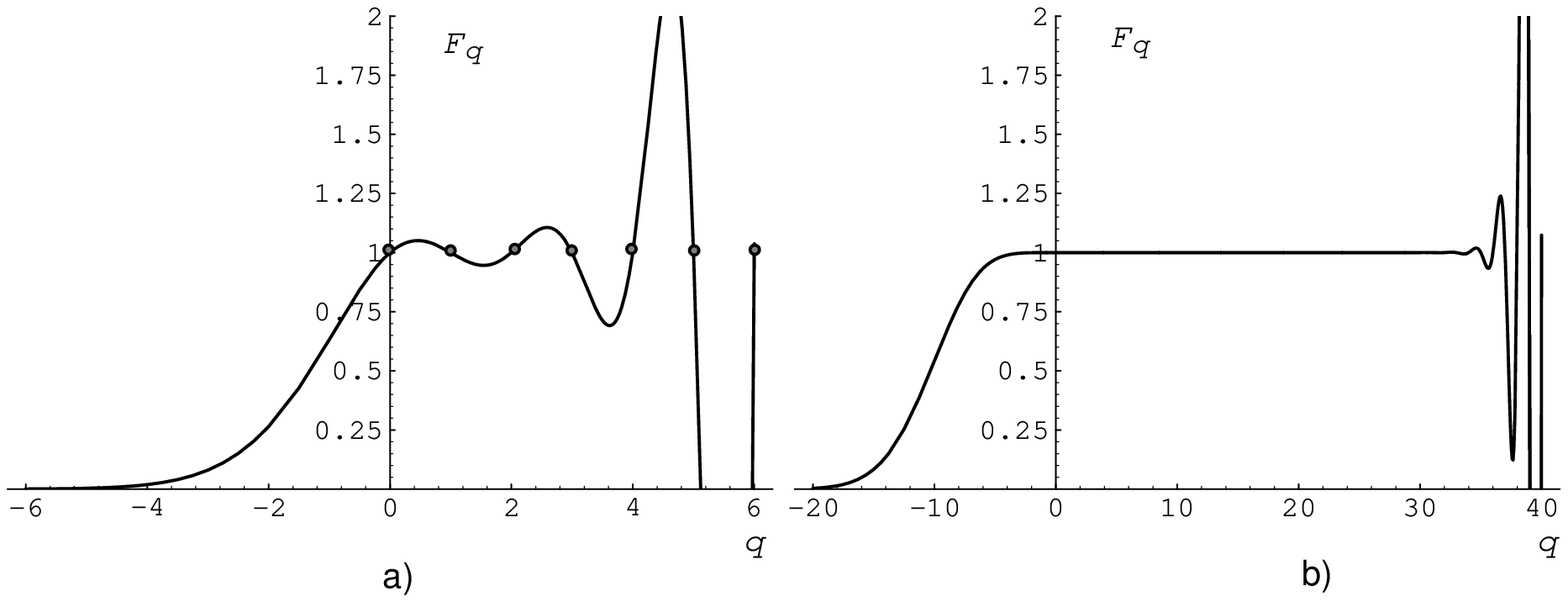}
\caption{The Poisson GFM; a) : $n_b$=1, b) : $n_b$=10}
\label{poisson}
\efg
\brry
F_q&\eq&n_b^q{\gamma(-q,n_b)\over\Gamma(-q)}\rc
\clFq&\eq&F_q/n_b^q\eq\gamma^*(-q,n_b)
\erry
where $\gamma(-q,n_b)$ is the incomplete $\gamma$ function and  
$\gamma^*$ the analytical incomplete $\gamma$ fun\-ction\cite{Grad}.  
In any case, the value of $\gamma^*$ on the integer and positive (or  
0) values of $q$ is 1 which is natural since one recovers here the  
standard factorial moments of the Poisson distribution. But one has  
to
notice that, besides those points, the shape and behaviour of this  
function depends drastically on the $n_b$ value; This function has in  
fact two types of behaviour. One for $n_b\simeq 1$ and one for $n_b  
>> 1$. This is illustrated in figure \ref{poisson}.

If $n_b\eq1$ say, figure (\ref{poisson} a) exhibits a steeply  
oscillating behaviour for $q\geq0$ and goes rapidly to 0 when $q$ is  
negative. This behaviour could prevent us from using those moments  
for small values of $n_b$ where one cannot wait for a faithful  
behaviour of the moments and where the numerical formula (\ref{Gdef})  
can be very unstable. This is not a surprise if one considers that  
those moments are devoted to the study of rare events of low n, ie  
$n<<n_b$
For $n_b>>1$, say $n_b=10$, figure (\ref{poisson} b) shows that the  
$\gamma^*$ function is practically 1 in a large interval of $q$ :  
$-n_b<<q<<3 n_b$.

>From the point of view of the intermittency data, this indicates  
that the Poisson noise will be disentangled\cite{Generev} from the  
dynamical signal only for those $q$ greater than $-n_b$ . As in those  
data the mean value of the number of particle in each bin tends  
rapidly to 0, one can understand that the fractional part of the  
moments does not gives any dynamical insight on the basic process.
\bfg{thb}
\fg{15}{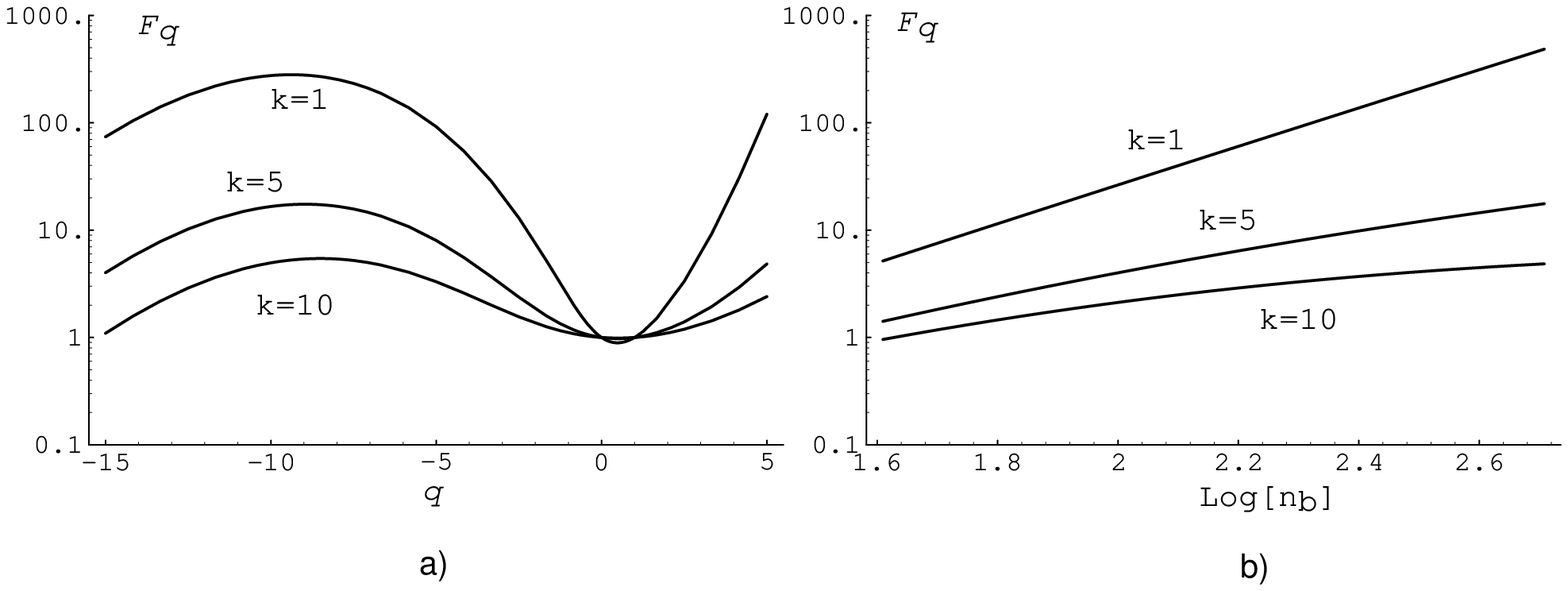}
\caption{The GFM of the Negative Binomial Distribution. a) $q$  
behaviour for $n_b=10$; upper curve $k=1$, intermediate one : $k=5$,  
lower one : $k=10$; b) $n_b$ behaviour at fixed $q=-5$, upper curve  
$k=1$, intermediate one : $k=5$, lower one : $k=4$}
\label{nbd}
\efg
\paragraph{\it ii) Self similar distributions.} The best way to buid  
an asymptotic self-similar distribution is to  construct $P_n$ as a  
compound Poisson distribution (a particular case of these  
distributions is the NBD distribution). At the level of the  
generating function, this gives:
\beq
G(z)\eq H(z-1),\ \ H(u)\eq\int_0^\infty \varphi(x)\expp{ux}dx,
\eeq
where $\varphi$ is the usual KNO function :
\beq
 P_n {\simeq}\lim_{n_b\to \infty}
\varphi(n/n_b)/n_b,
\eeq
and one gets:
\brry
\label{ssq}
H(u)& \eq &h(u n_b)\rc
F_q&\eq&{n_b^q\over \Gamma(-q)}\int_0^{n_b} h(-u)u^{-q-1}du
\erry
On this integral, one can notice that if $n_b$ is sufficently large,  
$\clFq =F_q/n_b^q$ tends to a constant and we recover the KNO result  
provides $h(u)u^{-q} \to 0$ when $u\to \infty$. Unless this condition  
is fulfilled, $\clFq$ will depends on $n_b$.

If, say, $h(-u)\simeq u^\alpha$, when $u\to \infty$, $\clFq$ will be  
KNO for $q>>-\alpha$. When $q<-\alpha$, the integral in \ref{ssq}  
will be dominated by the high u behaviour of the integrand and
\beq
\label{nbdev}
\clFq \ \simeq\ n_b^{-q+\alpha}
\eeq
Measuring the generalised moments of the distribution provides  a  
rather nice tool for the study of the high $u$ behaviour of the  
generating function.

As an exemple of this situation, we have calculated the moments of  
the Negative Binomial distribution~:
\beq
h(u)  \eq  \frac{1}{( 1-u/k)^k}
\eeq
which gives~:
\beq
\clFq \eq \frac{_2F_1(k,-q,1-q,-n_b/ k)}{ n_b^q \Gamma(1-q)}
\eeq
where $_2F_1$ is the hypergeometric function of the second  
kind\cite{Grad}.

The general trend of the reduced generalised moments of the  
distribution is given in figure (\ref{nbd}-a) while the power-like  
behaviour of the moments is shown in fig (\ref{nbd}-b).
%%%%%%%%%%%%%%%%%%%%%%%%%%%%%%%%%%%%%%%%%%%%%%%%%%%%%%%%%%%%%%%%%%%%%% 
%%%%%
\section{QCD Generalised Moments.}
\subsection{ QCD Double Leading-log Approximation.}
The QCD evolution equation for the generating function $G(Q,z)$ of  
the multiplicity distribution produced by a parton of energy E  
disintegrating in a cone of opening $\Theta_0$ has been calculated  
since a while in the DLA approximation\cite{QCD} and reads~:
\beq
\label{qcdg}
\frac{\partial G(Q,z)}{\partial \log(Q)}\eq G(Q,z)\int_0^1 \go^2  
\frac{(G(xQ,z)-1)dx}{x}
\eeq
where $Q=E\Theta_0$ is the hardness scale. Notice that an implicit  
infrared cut-off must be understood in this equation, $Q x > Q_0$.  
This cut-off tends to 0 in the weak coupling regime of the equation  
and will be of importance in section {3-2}.

Let us  first fix some notations. As in the preceding section, we  
define $H(Q,u)=G(Q,1-u)$, and the self similar solution (KNO) of the  
equation \ref{qcdg}, $ h(v) $, such as $H(Q,u)=h(n_b u)$. Further,  
for negative u, let us define $\h(y)=h(-\expp{y})$.

With these notations the QCD solution obeys an integro-differential  
equation which does not depends explicitely on the coupling constant  
$\go$, and reads (using ref.4 with some slight change of notations)~:
\beq
\label{qcd-kno}
{d^2 \log(\h)\over dy^2}\eq\h-1
\eeq
This equation has been solved in an implicit way (for negative  
$u=-\expp{y}$\cite{QCD}) :
\brry
y-y_+ &\eq& \int_{\log (2)}^{X(y)}{du \over  
\sqrt{2(u-1+\expp{-u})}}\rc
X(y)&\eq&\log(1/\h(y))\ , \ \ \ y_+=-.251
\erry
Starting from formula (\ref{ssq}), the QCD GFM for negative $q$ can  
be given as an $y$ integral:
\beq
\label{qcdfq}
\clFq\eq \frac{1}{\Gamma(-q)}\int_{-\infty}^{\log  
(n_b)}\expp{-X(y)-qy}dy
\eeq
This expression is well adapted to the steepest descent technique  
which gives:
\brry
\label{stds}
&-X(y)-q y\simeq -(y-y^*)/2\sigma^2 -1,\rc
&y*\eq -q-c_0,\ \ \sigma^2=1+\expp{-q^2/2-1},\ \ c_0=.41
\erry
where $c_0$ has been numerically computed for $y>2$.
This gives asymptotically ($n_b\to \infty$):
\brry
\log(\clFq^{qcd})&\eq& q^2/2-(1-c_0)  
q-1+(q-.5)\log(-q)+\log(\sigma^2)/2,\rc
q&<<&-2
\erry
where we use the same approximation (steepest descent) for the  
$\Gamma$ function.
\bfg{thb}
\fg{12}{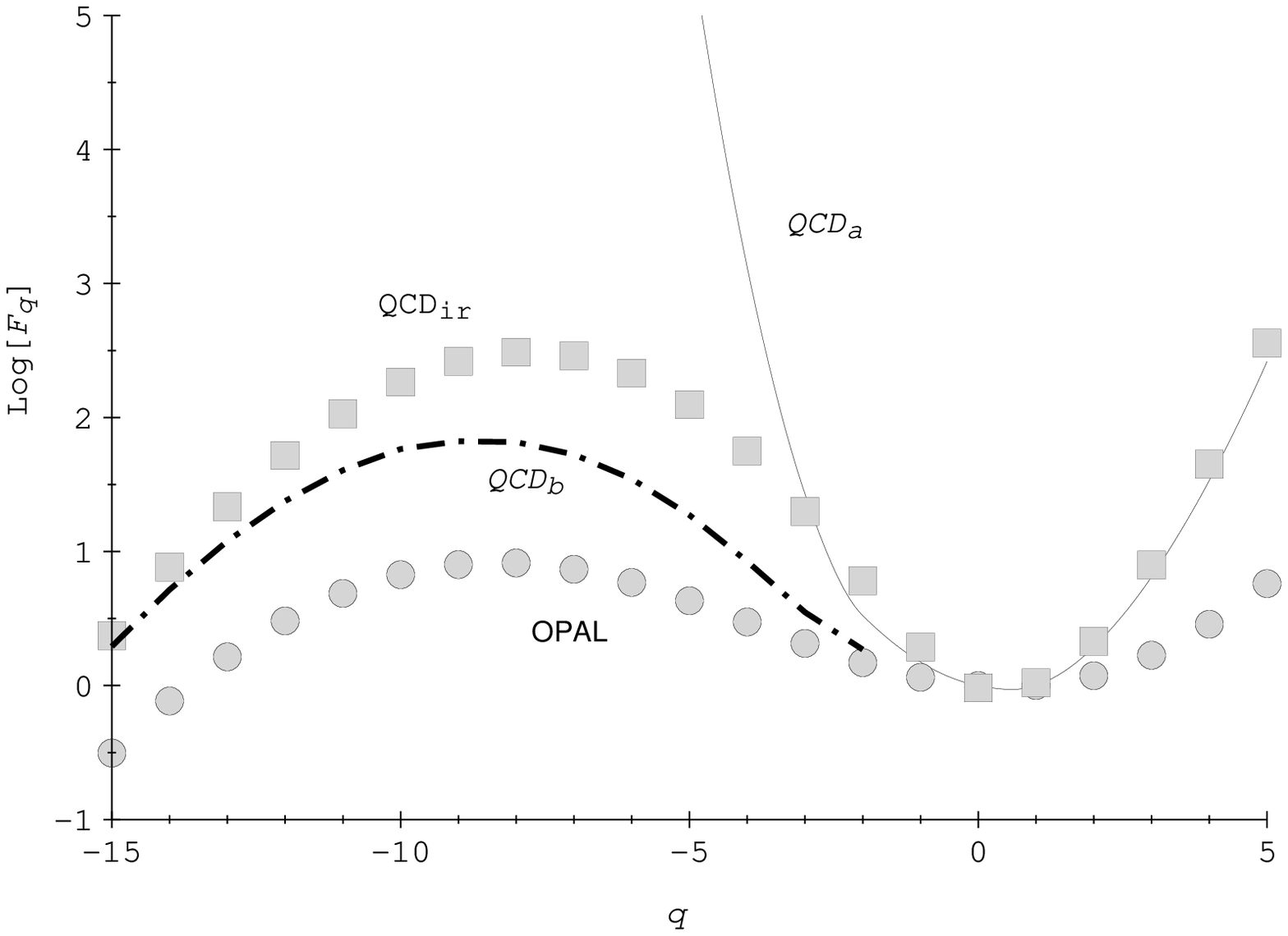}
\caption{GFM for negative values of $q$. Continuous line : Asymptotic  
QCD; dotted-dashed line : finite $n_b$ corrections to asymptotic  
theory; squares :  finite infrared corrections (Monte-Carlo); rounds  
: OPAL data (one hemisphere multiplicity)}
\label{Fqcd}
\efg
One has now to deal with finite $n_b$ corrections to this formula.  
First we verify that the Gaussian approximation is a lower bound for  
$\h$, the same for the the GFM calculated with formula \ref{qcdfq}.  
Using formula \ref{qcdfq} with the Gaussian approximation for $\h(y)$  
as a lower bound for the finite size QCD GFM gives :
\beq
\label{qcdna}
\clFq(n_b)\eq\clFq^{qcd} \Biggl\{ {1-{\rm  
erf}[-(\log(n_b)+c_0+q)/\sqrt{2\sigma^2}] \over 2 } \Biggr\}
\eeq
The theoretical prediction are shown in figure \ref{Fqcd} together  
with the experimental data from OPAL\cite{OPAL}. We have taken the  
multiplicity data on one hemisphere in order to be as close as  
possible to the one parton multiplicity distribution.
Notice that the prediction of the theory including finite $n_b$  
effects has to  be understood as a lower bound for the moments,  
specially for small values of $-q$.

\subsection{Infrared effects.} Let us now turn back to formula  
(\ref{qcdg}). We have now to take into account the finite infrared  
cut-off, $Q_0$ of the theory.

In order to be coherent with the probabilistic interpretation of QCD  
DLA, the decay probability, $\Pi(x)$ of a parton of hardness $Q$ into  
two partons of hardness $xQ$ and $(1-x)Q$, must be normalised  
positive and definite . Keeping tracks only for the logarithmically  
divergent part of the Altarelli Parisi kernel, we have~:
$$
\Pi(z)=\delta(z)+\frac{\go^2}{z}\Big|_+\ =\lim_{\epsilon\to  
0}\Biggl\{\frac{\go^2\Theta(z-\epsilon)}{z}+\delta(z)(1-\go^2\log(1/\epsilon))\Biggr\}
$$
where $\Theta(x)$ is the Heaviside step function.

In the fixed (and finite) coupling constant case, one can observe  
that the probability distribution is no more positive definite in the  
limit $\epsilon \to 0$. As a consequance, this limit is incompatible  
with the building of a  Monte-Carlo calculation. This is a reflection  
of the fact that the perturbative theory is exact only in the limit  
where $\go\to 0$, i.e. at infinite energy.

To evade this difficulty, one has to use a finite cut-off $Q_0$. The  
simplest choice is given by $\log(Q/Q_0) \simeq 1/\go^2$  wich gives  
$\Pi(z)=\go^2\Theta(z-\expp{1/\go^2})/z$. Using this form in equation  
(\ref{qcdg}) gives at the level of the self-similar solution $\h$~:
\bfg{thb}
\fg{12}{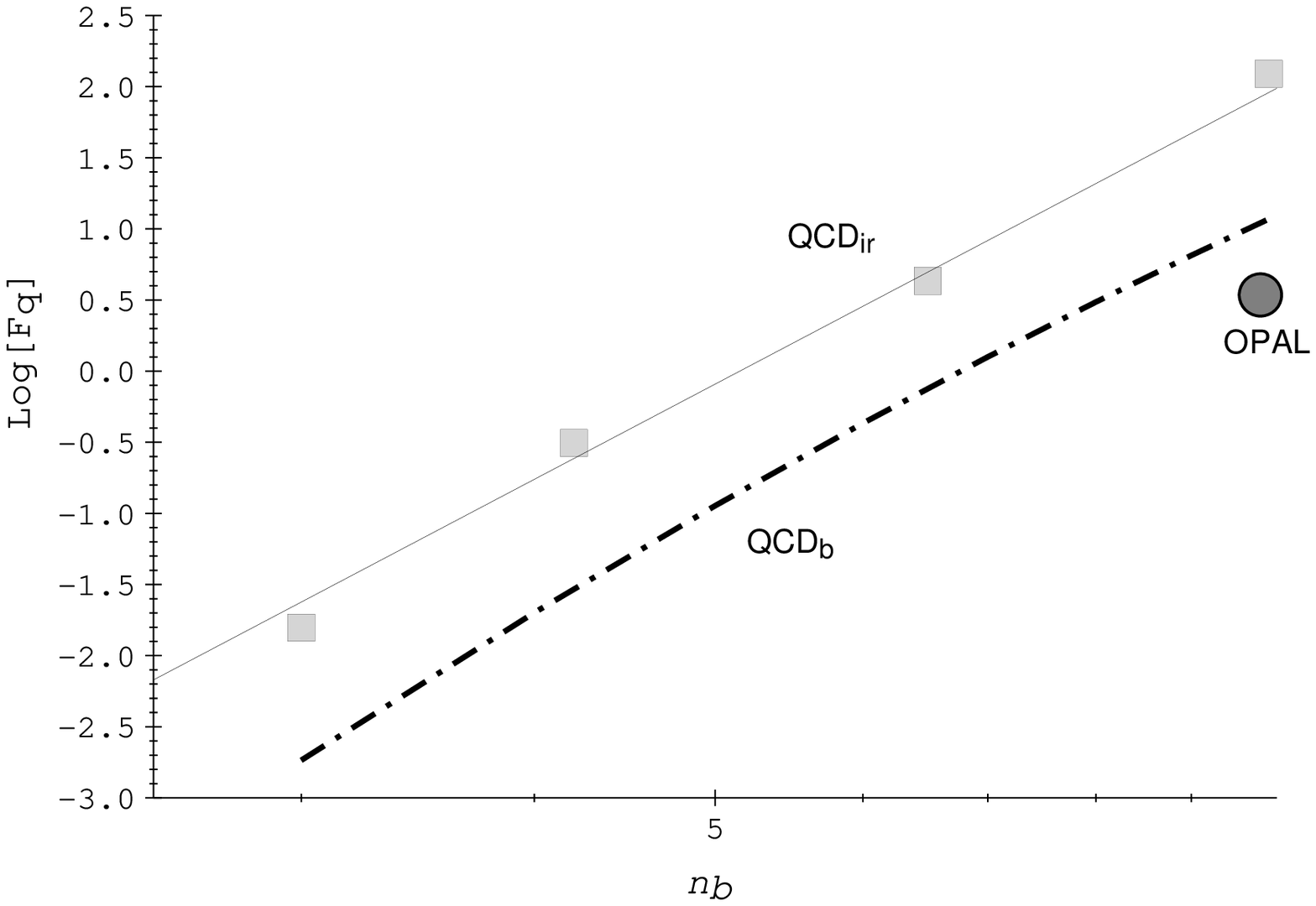}
\caption{ $n_b$ behaviour of ${\cal F}_{-5}$ : Continuous curve  
${\cal F}_{-5}\propto n_b^{3}$, grey squares : QCD Monte-Carlo,  
dotted-dashed one QCD lower bound for finite size effects.}
\label{fqnb}
\efg
\beq
\label{irsse}
\frac{\h'(y)}{\h(y)}\eq\int_{y-1/\go}^y \h(z) dz -\frac{1}{\go}
\eeq
which has an asymptotic solution :
\beq
\label{iras}
\h(y)\ \propto\ \expp{-\frac{y}{\go} },\ \ \ y>>\frac{1}{\go}
\eeq
in the limit where $y$ goes to infinity.

This exponential behaviour is different from the Gaussian solution of  
the QCD case (\ref{stds}). For finite $\go$,  the infrared cut-off  
have changed the asymptotic behaviour
of the self-similar solution of the QCD evolution equation. Notice  
however  that one recovers the previous solution when $\go$ goes to  
0.

 As the GFM of negative order are sensitive to the high y values of  
the generating function, it is not astonishing that they show a  
different energy behaviour~:
\beq
\label{fqas}
\clFq\ \propto\  
n_b^{-q-\frac{1}{\go}}\eq\big[\frac{E\Theta_0}{\mu}\big]^{-(1+\go q)}
\eeq
This behaviour is of the NBD type (see Eq. (\ref{nbdev})) is well  
reproduced by numerical (Monte-Carlo) calculations, even for low  
$n_b$.

Let us now discuss a little the QCD results. In figure \ref{Fqcd} the  
Monte-Carlo results are shown to be higher than the lower bound we  
derived in the weak coupling limit. We verify that this does not  
depends on the detail of the Monte-Carlo. The infrared structure of  
the theory, namely the cut-off in the evolution equation,  enhances  
the fluctuation pattern for negative values of $q$. Notice that the  
Monte-Carlo calculation gives the known results for positive values  
of $q$ which shows that the infrared cut-off has no particular effect  
on the standard positive moments. The OPAL data have been presented  
in this figure to show that the data (one hemisphere data) are  
substantifically lower than DLA  QCD predictions in all the  $q$  
range. Figure \ref{fqnb} shows the $n_b$ behaviour of the GFM of  
order -5. It is interesting no notice that while the two QCD results  
are different in strength, the apparent slope of the QCD bound is  
close to the Monte-Carlo one and is very close to the one predicted  
in formula (\ref{fqas})$(-q-1/\go\simeq 3 $ at LEP) .
\section{ Conclusions.} In this paper we have presented a  
comprehensive generalisation of the standard factorial  moments which  
have been shown to be sensitive to the infrared structure of the  
theory. This analysis focusses on the low multiplicity events of high  
energy reactions such as ${\cal Z}_0$ decay at LEP. The QCD GFM have  
been calculated together with a bound on low energy corrections and  
have been shown to be substancially higher than the OPAL data.  When  
one includs the natural  infrared cut-off in the theory, the  
asymptotic picture of the GFM are modified in the negative part of  
its spectrum and enhances the fluctuation pattern.    Our feeling is  
that, in contradiction  to the positive $q$ case where Next to  
Leading-Log corrections (energy conservation effects\cite{D93})  are  
probably enough to explain the discrepancy of the asymptotic theory  
with experiments, the GFM of negative order will emphasis the non  
perturbative part of the theory and could provide a glance on the  
hadronisation part of the strong interaction theory. Further work is  
in progress in this direction.

\paragraph{\bf Acknowledgements} It is pleasure for one of us (J.-L.  
M) to thanks R.Hwa and R. Peschanski for numerous and fruitful  
discussions on the subject, M. Le Bellac and, once more, R Peschanski  
for a careful reading of the manuscript.

\end{document}